# Suspended lithium niobate acoustic resonators with buried electrodes for radiofrequency filtering

Silvan Stettler [1], Luis Guillermo Villanueva [1, †]

[1] *Institute of Mechanical Engineering, École Polytechnique Fédérale de Lausanne (EPFL), 1015 Lausanne, Switzerland*

Data rates and volume for mobile communication are ever-increasing with the growing number of users and connected devices. With the deployment of 5G and 6G on the horizon, wireless communication is advancing to higher frequencies and larger bandwidths enabling higher speeds and throughput. Current micro-acoustic resonator technology, a key component in radiofrequency front end filters, is struggling to keep pace with these developments. This work presents a novel acoustic resonator architecture enabling multi-frequency, low-loss, and wideband filtering for the 5G and future 6G bands located above 3 GHz. Thanks to the exceptional performance of these resonators, filters for the 5G n77 and n79 bands are demonstrated, exhibiting fractional bandwidths of 13% and 25% respectively with low insertion loss of around 1 dB. With its unique frequency scalability and wideband capabilities, the reported architecture offers a promising option for filtering and multiplexing in future mobile devices.

The fifth and sixth generation of wireless communication technology, commonly referred to as 5G and 6G, promise unprecedented advancements in data transfer rates, network capacity, and connectivity. Within the spectrum allocated for 5G new radio (5G NR) deployment, the sub-6 GHz range (5G Frequency Range 1, 5G-FR1) stands out as a crucial spectrum range for large area coverage. The mid-band spectrum (1-6 GHz) strikes an optimal balance between capacity and coverage making it particularly suited for enabling high-speed mobile services in urban and suburban areas [1], [2]. One of the key factors for enhanced data rates compared to 4G Long Term Evolution (LTE) is the increased bandwidth available with the new 5G NR bands such as the n77 (3.3 – 4.2 GHz), n78 (3.3 – 3.8 GHz), and n79 (4.4 – 5.0 GHz) bands featuring fractional bandwidths (FBW) as high as 24 % (n77). The combination of higher frequencies and expanded bandwidths results in increasingly demanding requirements for filter hardware in the RF Front End (RFFE) based on surface acoustic wave (SAW) and bulk acoustic wave (BAW) resonators [3]. At the resonator level, up-scaling resonance frequencies ($f_r$) requires further miniaturization of the acoustic wavelength ($\lambda$) and high acoustic velocities ($v_p$), while maintaining low electrical and acoustic losses. Simultaneously, addressing bands with large FBW with acoustic filters necessitates correspondingly high effective electromechanical coupling ($k_{eff}^2$) of the individual resonator building blocks.

These challenges have motivated advances in incumbent SAW and BAW technologies in the past decade.

---

† guillermo.villanueva@epfl.ch

Replacing aluminum nitride (AlN) with scandium-doped aluminum nitride (AlScN) has boosted $k_{eff}^2$ for BAW devices due to the larger $d_{33}$ coefficient of AlScN [3], [4]. Even though BAW filter technology has been successful in the 1-8 GHz range [3], [5], [6], [7], the $k_{eff}^2$ of AlScN is still not large enough for the broad n77-79 bands. SAW resonators based on lithium niobate (LiNbO₃) or lithium tantalate (LiTaO₃) can achieve higher $k_{eff}^2$ and $f_r$ can be adjusted by lithography changing the electrode pitch of the interdigitated transducer (IDT) [8]. Using multi-layered substrates, such as thin films of LiNbO₃ or LiTaO₃ bonded to high-velocity materials [9], [10], [11], [12], [13], [14], [15] or Bragg reflector stacks [16], has improved $k_{eff}^2$, quality factor ($Q$) and enabled scalability to 3 GHz and beyond. Wideband, low insertion loss (IL) filters for the n77-79 bands have been demonstrated recently using these LiNbO3-SiC [14], [17], [18] and LiNbO3-SiO2-Si [19] heterostructure SAW resonators.

Resonators exploiting acoustic waves in suspended films of LiNbO₃ or AlScN such as symmetric [20], [21], [22], [23], [24] and antisymmetric [25], [26], [27] Lamb waves, and shear horizontal (SH) waves [28], [29], [30], have garnered interest as alternatives to classical BAW and SAW resonators. Since the deformation of the piezoelectric layer is not constrained in this configuration, plate waves generally demonstrate higher $k_{eff}^2$ than what is possible with SAW modes, which facilitates synthesis of wide bandwidth filters. The suspended structure also reduces leakage of acoustic energy, resulting in low substrate losses even at high frequencies [31]. So far, suspended LiNbO₃ resonators operating with the first-order antisymmetric Lamb mode (A1, i.e. XBAR) [25], [32], [33], [34] and first-order shear horizontal mode (SH1) [35], [36] have shown promise for wideband filtering in the 5G mid-band spectrum owing to the high $v_p$ and $k_{eff}^2$ of these modes. However, the achievable $f_r$ in these types of devices is strongly dependent on film thickness and lithographic tunability is limited. *Propagating waves* such as the fundamental shear horizontal (SH0) and symmetric (S0) modes in suspended LiNbO₃ films provide greater flexibility. Like SAWs, SH0 and S0 modes can be excited with IDTs and thus benefit from lithographically defined $f_r$. Furthermore, $v_p$ of these modes is only weakly dependent on the film thickness, allowing for a wide range of $f_r$ on a single substrate. Although SH0 and S0 resonators with high $k_{eff}^2$ have been reported [28], [30], [37] around 1 GHz, scaling to 5G mid-band frequencies has been difficult. Further decreasing $\lambda$ and IDT pitch comes with a drastic drop in $k_{eff}^2$ which would impact filter bandwidth [20], [29], [37]. Reducing the thickness of the LiNbO₃ film and IDT electrodes can mitigate this tradeoff to a certain extent [29], [38], [39], but this presents practical challenges, including mechanical stability issues and increased electrical resistance.

In this paper, design and fabrication of an improved suspended LiNbO₃ resonator architecture that overcomes the frequency limitations of the SH0 and S0 modes is presented. The key novelty of the proposed resonator is the integration of buried IDT (B-IDT) electrodes (buried in the LiNbO3 layer) [40], [41], [42], instead of conventional IDT electrodes (S-IDT) (deposited on top of the suspended LiNbO₃ layer). This design achieves high $k_{eff}^2$ SH0 and S0 resonances at small $\lambda$ without the need to reduce LiNbO₃ film and electrode thickness. B-IDT resonators operating in SH0 and S0 mode co-fabricated on a 300 nm-thick YX36°-cut LiNbO₃-on-Si substrate are demonstrated in the 1-7 GHz range. The resonators exhibit high $k_{eff}^2$ over the whole frequency range reaching 32% at 3.5 GHz (SH0) and 15% at 5.8 GHz (S0) with $Q$ on the order of 100. With these advancements, purely acoustic filters for the full n77 and n79 bands using SH0/S0 resonators in a suspended configuration are achieved for the first time. The measured responses (with 50 Ω port impedances) show center frequencies of 3810 and 4780 MHz, 3dB FBW of 25% and 13%, and minimum IL of 0.8 and 1.5 dB, respectively. The reported results highlight the suitability of SH0 and S0 resonators for 5G filtering applications as an alternative to established XBAR-type suspended devices.

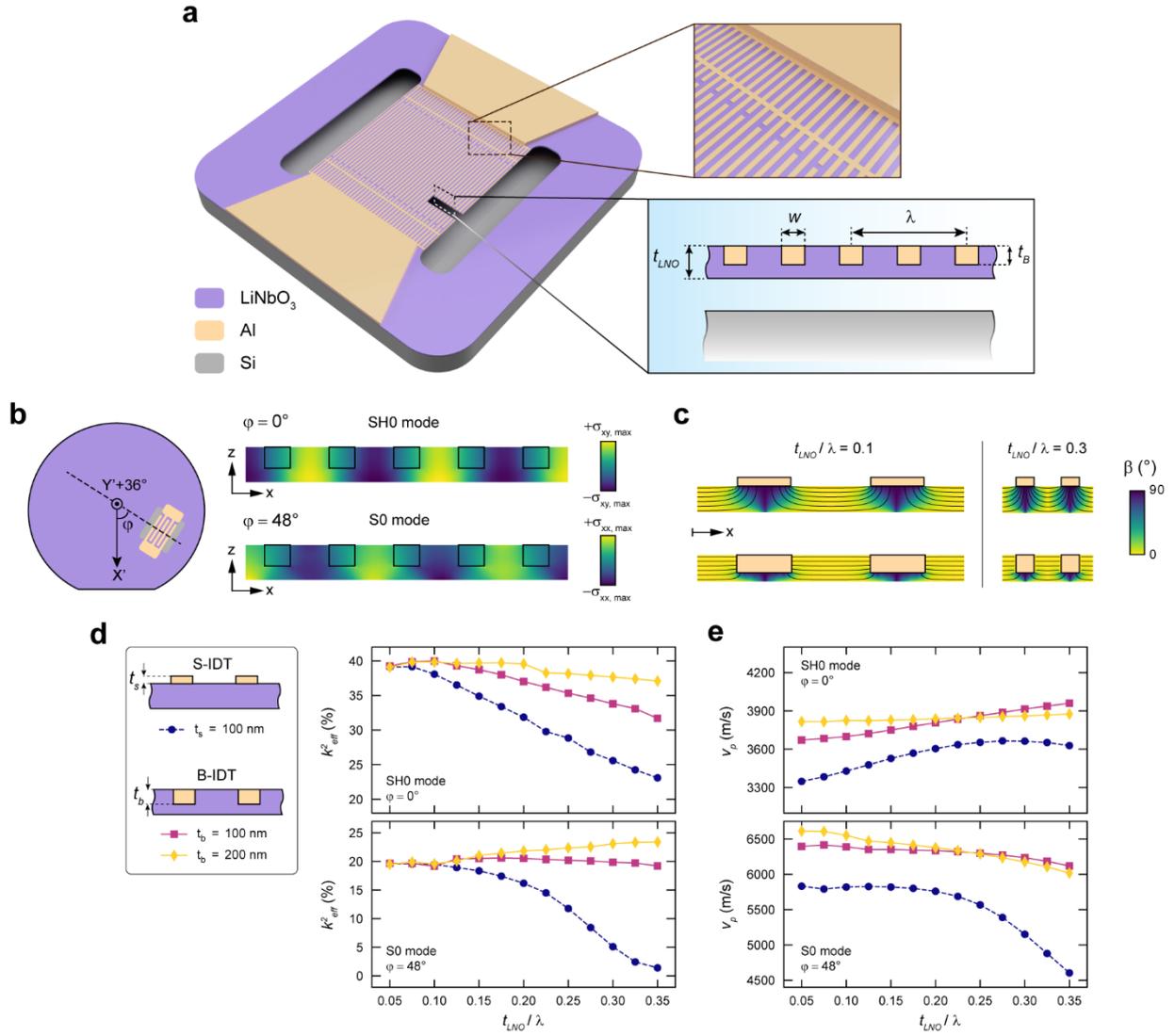

**Fig. 1 Concept and design of suspended plate resonators with buried IDT (B-IDT). a**, Schematic 3D view of a one-port resonator consisting of a suspended lithium niobate film with an array of B-IDT electrodes and cross-sectional view of the B-IDT. **b,** Orientation of the transducer on the wafer with respect to the crystalline X-axis of LiNbO₃ (denoted X') and corresponding stress distributions of the excited modes. **c,** Electrostatic simulations of the electric field in S-IDT and B-IDT transducer unit cells. Color shows the angle of the electric field with the horizontal direction ($\beta$). **d, e** Simulated $k_{eff}^2$ and $v_p$ of the SH0 and S0 modes as a function of $t_{LNO}/\lambda$ for S-IDT and B-IDT architectures with $t_{LNO}$ = 300 nm.

# Buried electrode resonator design and concept

Fig. 1a depicts an overview of the proposed suspended B-IDT resonator with a suspended film of LiNbO$_3$ with an array of electrode fingers with alternating polarity forming the IDT. The IDT is connected to regions with a thick metal layer forming the contact pads. The lateral boundaries are defined with an opening in the LiNbO$_3$ film. The free edge parallel to the electrodes acts as an edge reflector confining the acoustic energy in the suspended film [29], [37]. The defining characteristic of a B-IDT resonator is that the metal electrodes forming the IDT are embedded in the piezoelectric film rather than being deposited on top of it. For this demonstration, aluminum is used as an electrode material for its low density, high conductivity, and ease of use for microfabrication. A film of YX36°-cut LiNbO$_3$ is chosen as both the SH0 and S0 modes can be excited selectively by varying device orientation with respect to the crystallographic axis. If the exciting electric field is aligned with the crystalline X-axis, a strong SH0 wave is excited due to a large $e_{16}$ coefficient. At a 48° offset from the crystalline X-axis, $e_{16}$ is zero but $e_{11}$ is large instead, which enables excitation of the S0 mode (Fig. 1b) *without* simultaneously exciting the SH0 mode (see Supplementary Information, Section 1). To maximize $k_{eff}^2$ of these acoustic modes, it is crucial to ensure that the field between two adjacent electrodes is as horizontal as possible (parallel to the film surface) [31]. Fig. 1c shows the simulated electric field in S-IDT and B-IDT structures. For conventional S-IDT, the electric field is perfectly horizontal only near in the middle between two electrodes and the top surface of the film. Under the electrodes, the electric field is primarily vertical and does not contribute to the generation of the intended SH0 and S0 waves. In addition, vertical E-field components may excite undesired spurious modes due to non-zero $e_{3j}$ components in the piezoelectric tensor of LiNbO$_3$. These two effects are exacerbated when the ratio of piezoelectric film thickness ($t_{LNO}$) to $\lambda$ becomes large which is inevitably the case when upscaling $f_r$, unless $t_{LNO}$ is reduced. With B-IDT electrodes on the other hand, the electric field is always horizontal between the electrodes and areas with vertical components are strongly minimized. If the buried depth ($t_b$) is a significant fraction of $t_{LNO}$, such an optimal electric field distribution can be ensured even for high $t_{LNO}/\lambda$.

Fig. 1d shows the simulated evolution of $k_{eff}^2$ of the SH0 and S0 modes with $t_{LNO}/\lambda$ for the B-IDT and S-IDT architecture with a typical electrode thickness for reference. The efficacy of the conventional S-IDT architecture for high-frequency operation is compromised due to a significant decline in $k_{eff}^2$ with small $\lambda$. Especially the S0 mode cannot be excited efficiently at $t_{LNO}/\lambda > 0.3$ due to the deteriorating effect of the electrodes on the surface and the sub-optimal electric field distribution. In contrast, the degradation of $k_{eff}^2$ is reduced with a B-IDT architecture and high $k_{eff}^2$ can be attained for both modes even at $t_{LNO}/\lambda > 0.3$. Interestingly, for S0 it is possible to observe an increase in the $k_{eff}^2$ for large $t_{LNO}/\lambda$. Simulated acoustic velocities ($v_p = f_r \cdot \lambda$) for both SH0 and S0 modes are presented in Fig. 1e. Forming the electrodes by replacing parts of the piezoelectric film with metal rather than depositing on the surface leads to larger $v_p$, contrary to what one could intuitively expect a priori. Indeed, when using B-IDT the mode shapes are modified, resulting in higher $k_{eff}^2$ and $v_p$, even for small $\lambda$ (see Supplementary Information, Section 2). Another interesting point visible in Fig. 1d-e is that the dispersion of $k_{eff}^2$ and $v_p$ as a function of $t_{LNO}/\lambda$ is strongly reduced, which facilitates the fabrication of devices with lithographically tuned resonance frequencies spanning a broad spectrum while maintaining consistent performance metrics. Moreover, the ability to selectively excite two different modes offers the possibility to implement filters with varying bandwidth and frequency specifications on the same substrate.

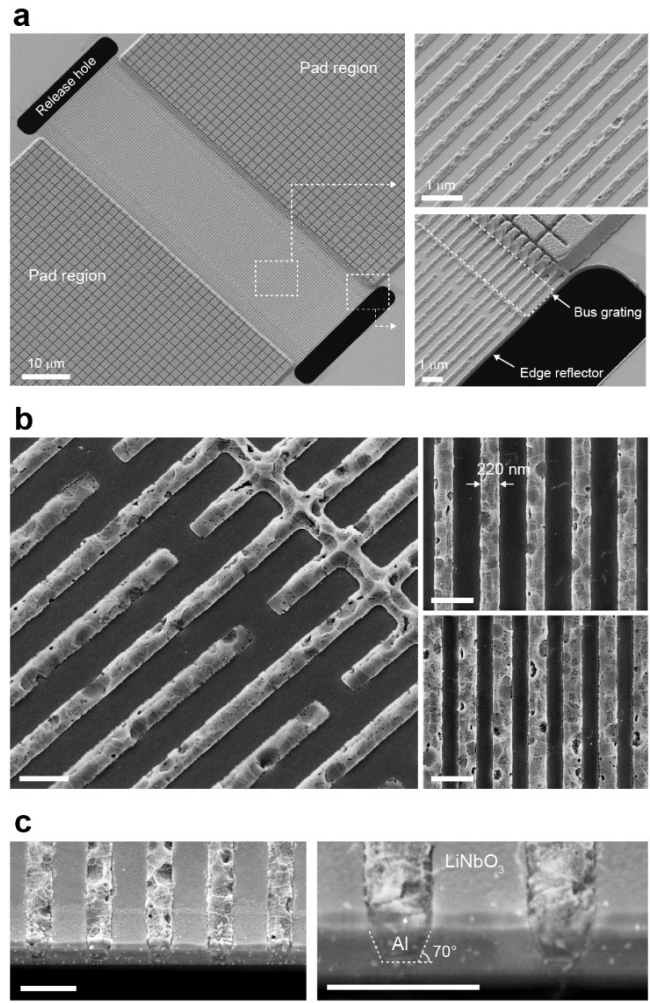

**Fig. 2 SEM micrographs of fabricated devices. a,** Overview of a fabricated device with suspended transducer, pad regions, and lateral release holes. **b,** Close-up views of the B-IDT. Scale bar 500 nm. **c,** Cross-sectional view of the B-IDT obtained by etching a slit through the LiNbO$_3$ film. Scale bar 500 nm.

Creating a structure with embedded metal features in the LiNbO$_3$ film presents several fabrication challenges including the patterning of narrow yet sufficiently deep trenches into the LiNbO$_3$ film followed by filling with Al leaving gaps between the metal and LiNbO$_3$ sidewalls. The optimized fabrication process flow for B-IDT resonators and filters is described in detail in the Methods. Fabrication is performed on chip level, but the process can easily be scaled to wafer level without any modifications. SEM micrographs showing an overview of a finished device are presented in Fig. 2a. A region with a thick Al layer that is monolithically connected to the B-IDT forms the contact pads. In Fig. 2b, close-up views of the surface of the B-IDT are shown. The developed polishing process does not damage the LiNbO₃ surface between the electrodes and can be controlled to adjust the height of the electrode protruding from the LiNbO₃. Devices with electrode pitches down to 425 nm ($\lambda$ = 850 nm) are fabricated while maintaining an electrode width of 220 nm (regardless of $\lambda$) for better process uniformity. Fig. 2c shows cross-sectional views of the IDT highlighting the geometry of the buried electrodes featuring sidewalls angled at 70° and a thickness of around 200 nm buried in the LiNbO₃ film ($t_{LNO}$ = 300 nm). 200 nm is chosen since this thickness is expected to yield favorable $k_{eff}^2$ and $v_p$ at $\lambda \leq 1\ \mu m$ (Fig. 1d-e) while still being feasible with the available processing capabilities.

## Resonator characterization

Fig. 3a shows the measured admittance of two SH0 B-IDT resonators operating in the 3-5 GHz range of the n77-n79 bands. The response exhibits strong SH0 resonances with $k_{eff}^2$ close to 30% and resonance and anti-resonance quality factors ($Q_r$, $Q_{ar}$) of over 100. Due to the high $k_{eff}^2$, around 700 MHz of separation between $f_r$ and the anti-resonance frequency ($f_{ar}$) with an impedance ratio of over three decades can be reached which is highly promising for large-bandwidth filters. Fig. 3b shows the measured admittance of two devices with the same layout as in Fig. 3a but rotated by 48° on the chip surface. Compared to the SH0 mode, the excited S0 mode has lower $k_{eff}^2$ but higher $v_p$ and thus allows frequencies higher than 5 GHz to be attained with the same $\lambda$. Furthermore, the reported performance metrics are reached with static capacitances ($C_0$) yielding impedances close to 50 Ω, suitable for the implementation of matched filters.

Similar to the well-known stopband phenomenon observed for SAWs in periodic gratings [43], a stopband also exists for SH0 and S0 waves propagating in the B-IDT grating (see Supplementary Information, Section 3). As shown in Fig. 3a, the small ripples in conductance below 3.2 GHz are due to propagating waves that exist below the lower stopband edge. The main resonance peak is observed at the upper stopband edge. For the SH0 devices in Fig. 3a, a prominent spurious mode appears seemingly in the stopband close to $f_r$. A similar mode, albeit less pronounced, is visible in Fig. 3b: a bump in the conductance at approximately 5.4 GHz for $\lambda$ = 1.1 $\mu m$. The appearance of these modes is linked to lithographic misalignment of the edge reflector with respect to the last IDT electrode. The misalignment is typically on the order of 100 nm or smaller and can vary across a chip. With $\lambda$ on the order of a micron, this is sufficient to significantly alter the reflection characteristics of the acoustic wave at the etched boundaries. Further discussion on this issue is given in Supplementary Information, Section 4. Importantly, this does not represent a major downside of these devices, since standard lithographic tools can now reach alignment accuracy down to 10nm.

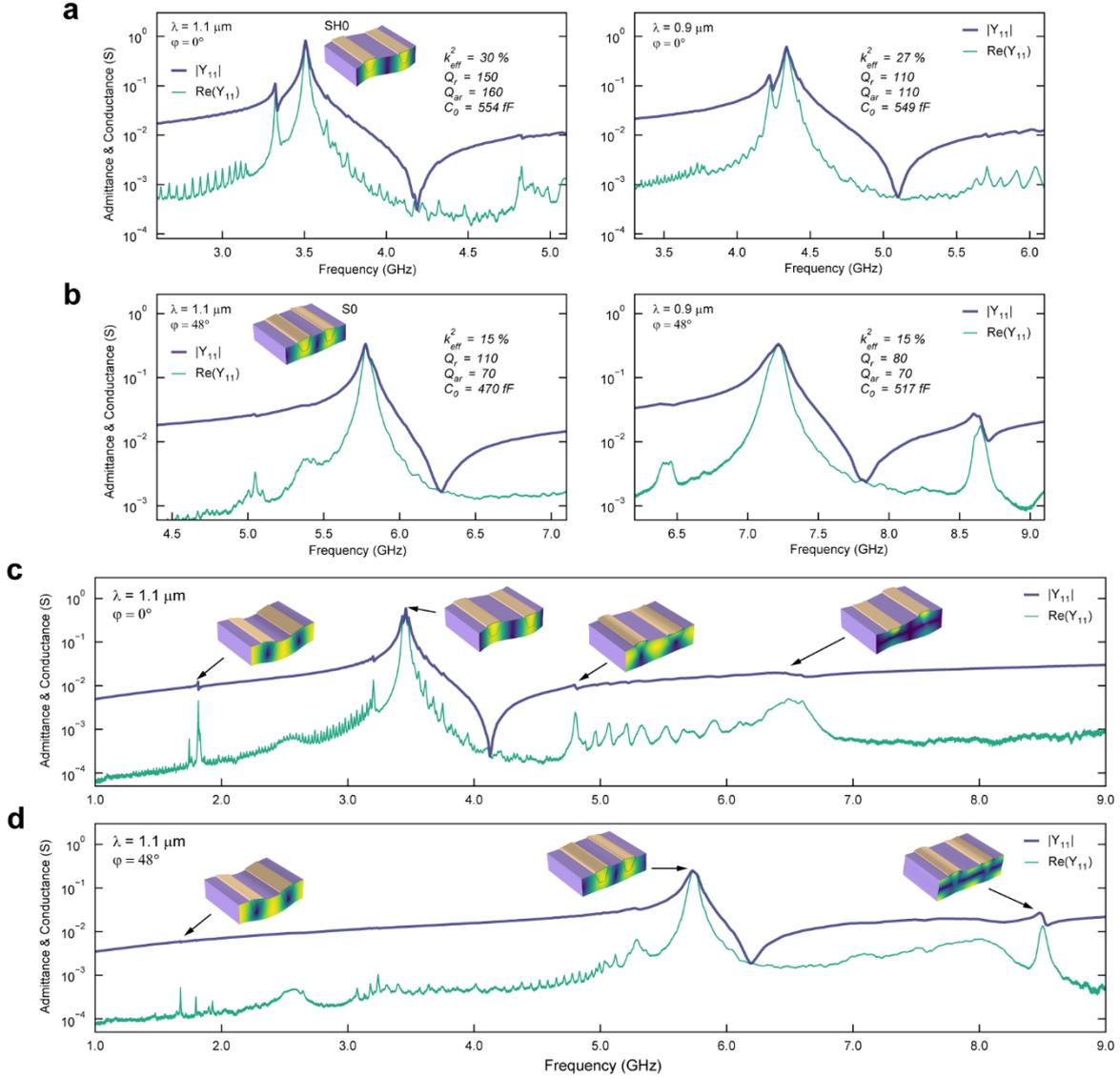

**Fig. 3 Measured admittance responses of buried IDT resonators. a, b,** Admittance of devices with $\lambda = 1.1$ μm, 0.9 μm operating in SH0 mode (a) and S0 mode (b). **c, d,** Admittance from 1-9 GHz of devices with $\lambda = 1.1\ \mu m$ operating in SH0 mode (c) and S0 mode (d). Inset mode shapes show the total displacement of resonating modes.

The admittance response over a wide frequency range for SH0 and S0 devices with $\lambda = 1.1\ \mu m$ is presented in Fig. 3c-d. Other than the small ripples generated by propagating waves, there are no dominant spurious modes up to at least 8 GHz. With the specific choice of on-chip device orientations, the SH0 and S0 modes never appear simultaneously in the same device. A small ripple below 2 GHz originating from the fundamental asymmetric Lamb mode (A0) remains due to the small but non-zero $e_{15}$ piezoelectric coefficient in YX36°-cut LiNbO$_3$. Since the applied electric field lacks vertical components, excitation of thickness-defined modes such as the S1 Lamb mode (Fig. 3c, 6.5 GHz) and Cross-sectional Lamé mode (Fig. 3d, 8.5 GHz) is very weak.

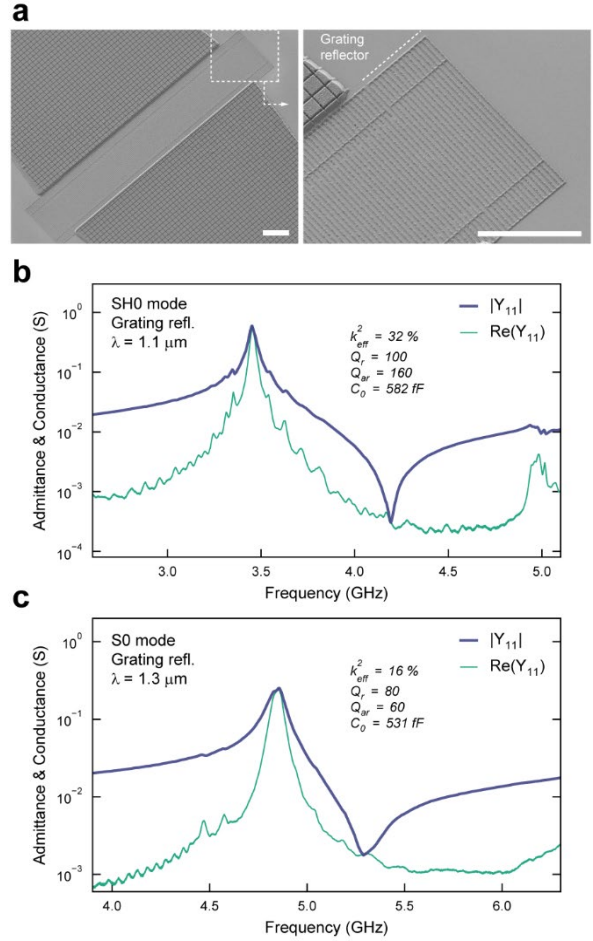

**Fig. 4 Measured admittance responses of buried IDT resonators with grating reflectors. a,** SEM images of a fabricated B-IDT resonator with grating reflectors consisting of a short-circuited grating of B-IDT electrodes. Scale bar 10 um. **b, c,** Admittance of B-IDT resonators with grating reflectors with λ = 1.1 um operating in SH0 mode (b) and with $\lambda$ = 1.3 $\mu$m operating in S0 mode (c).

To avoid the appearance of spurious modes in the vicinity of the main resonance due to misalignment of the edge reflectors, resonator designs with grating reflectors are implemented (Fig. 4a). Reflector and transducer elements are defined in the same lithographic mask, thus the issues of any misalignment of the reflection boundary are avoided. Fig. 4b-c show measured admittance responses of resonators with grating reflectors operating with the SH0 and S0 mode, respectively. The resonance frequencies do not change significantly compared to resonators with edge reflectors. The most notable advantage over resonators with edge reflectors is the absence of strong spurious modes to the left of the main resonance peak. The weak remaining ripples can be attributed to propagating modes outside of the B-IDT grating stopband and possibly transversal modes.

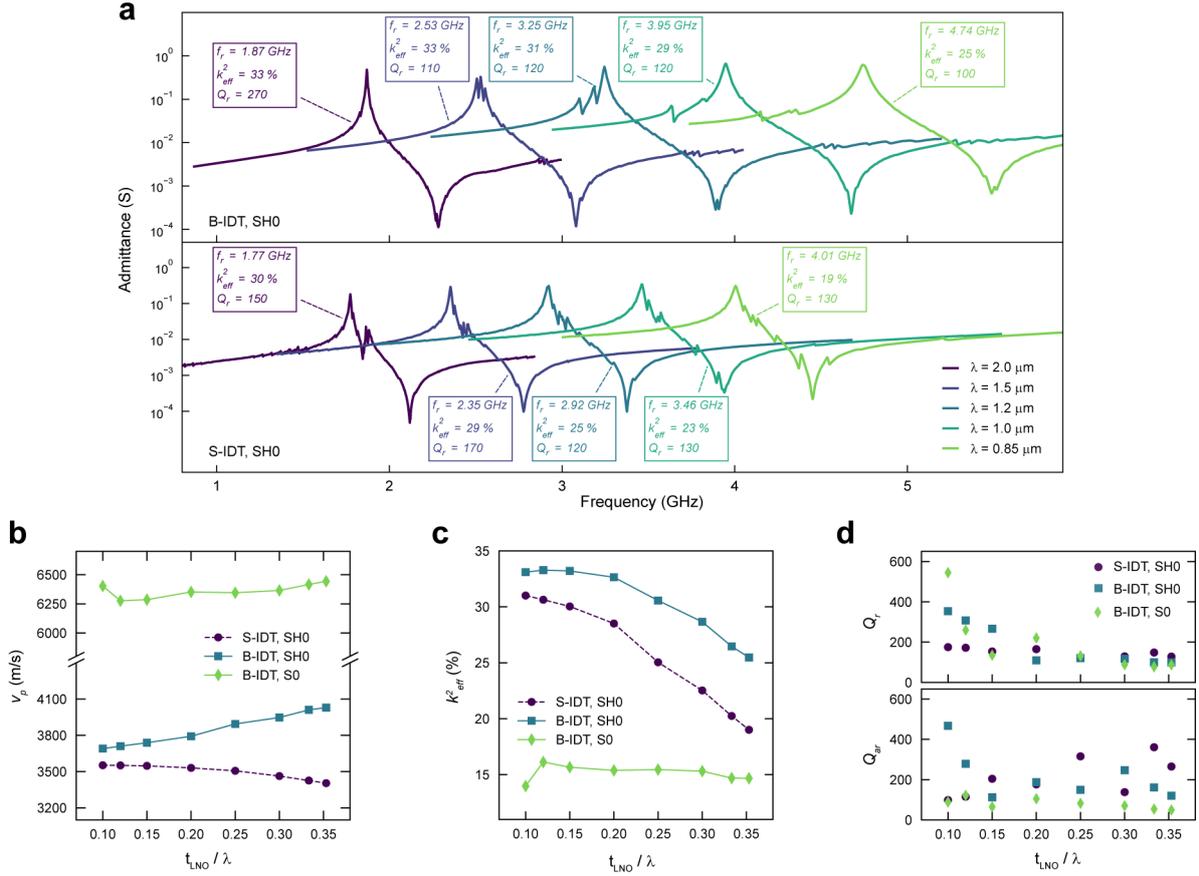

**Fig. 5 Measured admittance responses of resonators with different $\lambda$ and $t_{LNO}$ = 300 nm. a,** Admittance of B-IDT and S-IDT resonators with different $\lambda$ operating in SH0 mode. For a given $\lambda$, the transducer layout (number of electrodes, electrode length, and width) for compared B-IDT and S-IDT resonators is identical and employs edge reflectors. **b, c, d,** Summary of extracted $k_{eff}^2$ (b), acoustic phase velocity $v_p$ (c), and $Q_r$ and $Q_{ar}$ (d) for B-IDT and S-IDT resonators with edge reflectors.

To confirm the superior performance of the B-IDT over the conventional S-IDT architecture, a set of SH0 resonators with S-IDT gratings (Al electrodes, $t_s = 100$ nm) with $\lambda$ ranging from 850 nm to 3 μm is fabricated according to the process flow outlined in [37]. Fig. 5a compares measured admittance responses of B-IDT with S-IDT resonators with different $\lambda$. Extracted $v_p$ and $k_{eff}^2$ values are summarized in Fig. 5b-c. For devices operating in SH0 mode, it is evident that for the same distribution of $\lambda$, a broader range of frequencies can be covered with B-IDT resonators due to increased $v_p$ as predicted by simulations. At the same time, the drop in $k_{eff}^2$ is notably smaller for B-IDT resonators which further confirms their enhanced frequency scaling capability compared to S-IDT resonators. Remarkably, B-IDT devices operating with the S0 mode exhibit $v_p$ over 6000 m/s and $k_{eff}^2$ around 15% independent of the chosen acoustic wavelength. Fig. 5d shows the extracted $Q_r$ and $Q_{ar}$ for the fabricated range of $\lambda$. Towards higher $t_{LNO}/\lambda$ ratios, the values for $Q_r$ and $Q_{ar}$ for B-IDT resonators are decreasing and relatively similar which suggests acoustic losses are the dominating. While the B-IDT configuration allows for thick electrodes and thus low series resistance, a substantial portion of the resonating body consists of metal with generally much higher intrinsic acoustic losses than monocrystalline LiNbO$_3$ [31]. For filter implementation with low IL, resonator impedance at resonance typically needs to be on the order of 1 Ω or less. Despite the moderate quality factors (≈ 100) above 3 GHz, the B-IDT resonators reach impedances of 1 to 3 Ω at resonance owing to the enhanced $k_{eff}^2$. With the demonstrated $v_p$, $k_{eff}^2$, and $Q_r$ characteristics of the B-IDT resonators, the implementation of purely acoustic filters covering the full width of the n77 or n79 bands on the same chip can be envisioned.

## Wide-band filters for the 5G n77 and n79 bands

To demonstrate the suitability of B-IDT resonators for the 5G mid-band spectrum, filters for the n77 and n79 bands with a simple 2.5-stage ladder topology are synthesized and fabricated. For the n77 band filter (Filter 1), B-IDT resonators operating with the SH0 mode are chosen due to their superior $k_{eff}^2$, enabling an FBW of at least 24% to cover the entire band. For the n79 band filter (Filter 2) in the 4-5 GHz range, S0 mode B-IDT resonators are used, taking advantage of the high $v_p$ of the S0 mode. An optical microscope image of a fabricated filter is shown in Fig. 6a. Resonator designs with grating reflectors are opted for since the avoidance of spurious modes is critical for smooth filter pass bands. A circuit schematic and the acoustic wavelengths of individual resonators in the filters are provided in Fig. 6b. To ensure large out-of-band rejection, low IL, and sufficient matching to 50 Ω in the pass band, the capacitances of individual resonators are tuned by varying the aperture (electrode length) and the number of electrode pairs in the transducer. The measured S-parameter magnitudes of Filter 1 are shown in Fig. 6c. The transmission response features an ultra-wide passband with an FBW of 25% centered at 3.81 GHz enabled by the large $k_{eff}^2$ of the SH0 B-IDT resonators. Thanks to high $k_{eff}^2 \cdot Q_r$, the filter exhibits a small minimum passband IL of only 0.8 dB while simultaneously providing at least 25 dB of out-of-band (OOB) rejection in the lower stopband and at least 20 dB up to 7 GHz. The response of Filter 2 fabricated on the same chip is shown in Fig. 6d. The width of the passband corresponds to an FBW of 13% which is sufficient for the n79 band. The center frequency is slightly higher than required for the n79 band and could be shifted with small adjustments to the acoustic wavelengths of the resonators. The minimum passband IL is higher (1.5 dB) compared to Filter 1 due to the smaller $Q_r$ we obtain above 4 GHz. Filter 2 exhibits similar OOB rejection characteristics with at least 23 dB in the lower stopband and at least 20 dB up to 7 GHz. Despite the rather simple layout, containing five resonators with only three different acoustic wavelengths, the fabricated filters yield remarkable responses highlighting the exceptional performance of the B-IDT resonators.

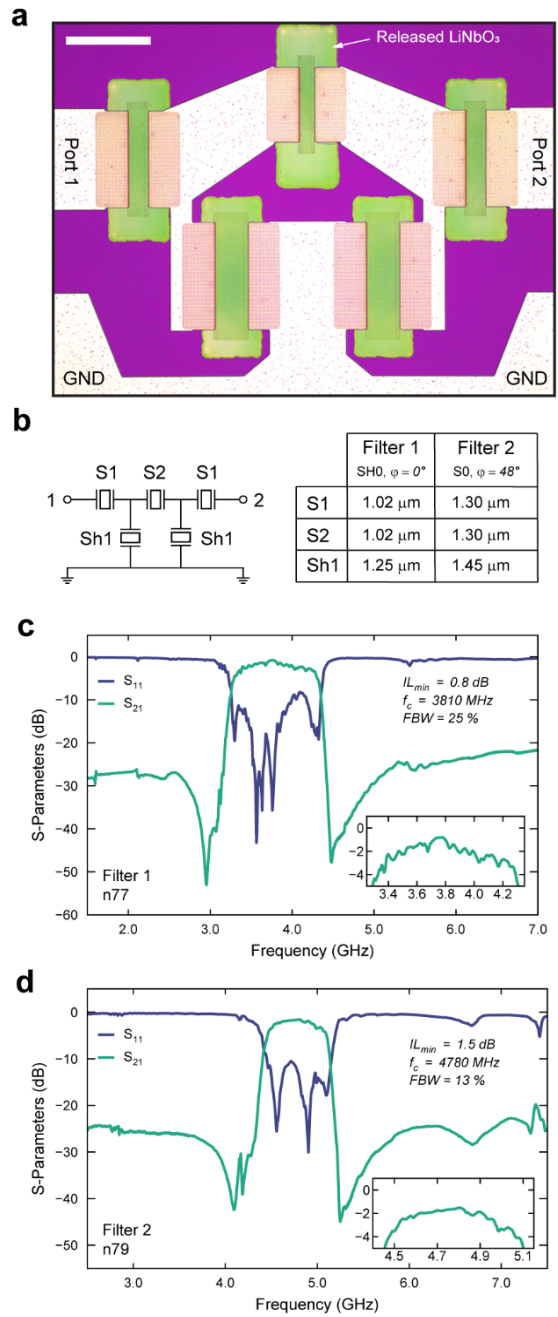

**Fig. 6 Design and measurements of filters using buried IDT resonators for the 5G n77 and n79 bands. a,** Optical microscope image of a fabricated filter. A design with grating reflectors is used for the resonators. Scale bar 100 um. **b,** Circuit diagram of the filter architecture and table listing $\varphi$ and $\lambda$ of the individual resonators. **c, d,** Measured S-parameters of a fabricated filter for the n77 band using resonators operating in SH0 mode (c) and for the n79 band using resonators operating in S0 mode (d).

# Conclusion

The reported resonator architecture, consisting of a suspended film of LiNbO$_3$ in combination with B-IDT electrodes, effectively enables the scalability of SH0 and S0 resonances to the 5G mid-band spectrum. The integration of B-IDT electrodes provides a solution to overcome the $\lambda$ versus $k_{eff}^2$ tradeoff that limits the potential of conventional S-IDT SH0 and S0 resonators for wideband filtering applications at high frequencies. The demonstrated resonator platform combines the advantages of lithographic modulation of $f_r$, orientation-dependent mode selectivity, and high $k_{eff}^2$ operation over an extensive frequency range. With these attributes, the integration of multiple filters for wide 5G bands with various center frequencies on the same substrate can be envisioned as highlighted by the presented filters for the n77 and n79 bands. In addition, it is feasible to implement filters composed of both SH0 and S0 resonators as building blocks for FBWs in the 13-24% range such as for the n78 band. Further improvements to the performance of the demonstrated n77 and n79 band filters such as passband flatness or OOB rejection could likely be achieved with more sophisticated filter designs and by increasing $Q_r$ and $Q_{ar}$.

At the resonator level, optimization of the fabrication processes for LiNbO$_3$ patterning and metal deposition selection and potentially an alternative electrode metal could be explored to minimize acoustic losses. Temperature compensation could be addressed by integrating a layer of silicon dioxide (SiO$_2$) between the electrodes or on the underside of the suspended film [30]. Moreover, considering future 6G deployment, further increases of $f_r$ up to 10 GHz or more without compromising $k_{eff}^2$ could be envisioned by adopting thinner LiNbO$_3$ film and smaller acoustic wavelengths in combination with B-IDT electrodes. Lastly, the B-IDT concept and presented fabrication method thereof could be applied to a variety of resonator types that rely on IDTs for transduction such as XBAR or SAW-type resonators.

## Methods

### Thin film LiNbO$_3$ substrates

For device fabrication, substrates consisting of a 300 nm thick, monocrystalline YX36°-cut LiNbO$_3$ thin film bonded to a high-resistive silicon wafer purchased from NGK Insulators are used. Prior to device processing, Pt alignment marks are deposited, and the wafers are diced into 14 by 17 mm chips.

### Fabrication process for B-IDT resonators

A schematic outline of the process flow is shown in Supplementary Information, Section 5. The outline of the process is inspired by the widely used Damascene process for interconnects [44] and previous demonstrations of SAW devices with buried electrodes [40], [41]. The most notable difference in the developed process flow is the absence of Chemical Mechanical Polishing (CMP) which is commonly used but prone to non-uniformities that may render a large portion of devices unusable across the wafer [41], [45]. Instead, an innovative polishing process is used that features excellent uniformity, material selectivity, and the possibility to be applied only locally. The process flow can be subdivided into four different parts. First, the trench pattern for the buried metal layer is etched into the LiNbO$_3$ film. Second, Al is deposited to fill the patterned trenches. Third, the surface is planarized with a polishing process. Finally, the release holes are patterned, and the device is released by removing the silicon underneath.

**Patterning of the LiNbO₃ film**

To start, a hard mask for LiNbO₃ etching consisting of a 150 nm Cr layer is deposited by e-beam evaporation. The Cr layer is patterned using e-beam lithography and ion beam etching (IBE, Veeco Nexus IBE350). Next, the underlying LiNbO₃ film is patterned by Reactive Ion etching (RIE) in an SPTS Advanced Plasma System (SPTS APS) etcher using CHF₃/Ar chemistry. The etch duration is controlled to achieve a trench depth of 200 nm. The layout of the etched features is designed to avoid aspect-ratio dependent non-uniformity of the etch rate (ER) and ensure that the trench depth is the same for all features. Specifically, all trenches for the electrodes are designed to have a width of 220 nm ($w = 200\ nm$) independent of $\lambda$. Further, any larger features such as the busbar and pad regions are replaced by dense gratings of trenches with the same width as the trenches for the electrodes. The profile of the etched structures features a flat bottom and a sloped sidewall (approx. 70°). The selectivity of this etch process with the Cr hard mask is 3:1. Thus, around 90 nm of Cr remains after LiNbO3 patterning which remains on the sample as it serves as a protective etch stop layer at a later stage of the fabrication process.

**Metallization of etched trenches**

The etched trenches are filled by evaporating a 10 nm Cr adhesion layer followed by 600 nm of Al. The sloped trench profile makes it possible to fill in the trenches with evaporation with good coverage of the sidewalls and without any voids in the metal [40]. Due to the directionality of evaporation, the topography of the features in the LiNbO₃ film remains in the deposited Al film.

**Local planarization with thick photoresist and dry etching**

At this point, the thick Al layer covers the entirety of the surface. This step aims to remove the excess Al that is not filling a trench or forming a contact pad.

The developed area-selective planarization process consists of planarizing the surface with a thick layer of photoresist (PR) followed by etch-back [46]. After Al deposition, a 700 nm layer of negative tone PR (AZ nLOF 2000, MicroChemicals GmbH, Germany) is spin-coated which fills in topographic features and creates a flat surface on top. The coated PR layer is subjected to a blanket exposure followed by a post-exposure bake to cross-link the entire film. Next, a second 700 nm thick layer of the same PR is spin-coated on top of the first layer and patterned using a maskless aligner (Heidelberg MLA150) to selectively increase the total PR thickness in areas dedicated to contact pads or filter interconnects. The etch-back step starts with IBE until the Cr layer covering the LiNbO₃ layer between the electrodes is reached. In areas covered with a double layer of PR, the underlying Al layer is not etched during this etch-back process. The protected Al thereby forms monolithic contact pads for the resonator or interconnects for the filters. In areas with only a single PR layer, Al only remains in trenches. However, the resulting surface topography in these areas is not necessarily flat and is dependent on the removal rates with IBE of the involved materials. Interestingly, crosslinked AZ nLOF resist etches slower than Al during the IBE etch-back process (selectivity AZ nLOF : Al $\approx$ 1 : 2), which leads to an inversion of the initial topography and thus excess Al on top of the filled trenches. These Al bumps are then flattened with RIE using $CHF_3$/Ar chemistry that features a high selectivity to Cr. The Cr layer protects the $LiNbO_3$ surface between the electrodes throughout all these etching processes. Finally, the remaining Cr is removed with wet chemistry that does not attack the buried Al electrodes (TechniEtch Cr01, MicroChemicals GmbH, Germany). Residual PR on the pad regions is stripped using $O_2$ plasma.

**Release hole patterning and device release**

The lateral release holes are patterned with E-Beam lithography using ZEP520A (Zeon Corp.) E-Beam resist and IBE. After stripping the resist mask, the devices are released by removing the underlying Si using $XeF_2$ gas.

**Device release for B-IDT resonators with grating reflectors**

Devices with a design including grating reflectors instead of edge reflectors are released by etching a cavity into Si from the back side of the substrate using Deep Reactive Ion Etching (DRIE) after the planarization step. Further details are given in [36].

## Device characterization and analysis

Individual resonators and filters are measured using a Rhode & Schwarz ZNB-20 Vector Network Analyzer (VNA) with an input power of -15 dBm. The admittance responses of the resonators are fitted with a modified Butterworth-van-Dyke (mBVD) model to extract $k_{eff}^2$, $Q_r$, $Q_{ar}$, $C_0$. $k_{eff}^2$ is defined as $(f_{ar}^2 - f_r^2)/f_{ar}^2$ according to [47].

# References

[1] D. Schnaufer, *5G RF For Dummies®, 2nd Qorvo Special Edition*. John Wiley & Sons, 2020.

[2] Nokia, "5G spectrum bands explained — low, mid and high band," 5G spectrum bands explained — low, mid and high band. Accessed: Jun. 14, 2024. [Online]. Available: https://www.nokia.com/thought-leadership/articles/spectrum-bands-5g-world/

[3] R. Aigner and G. Fattinger, "BAW Filters for 5G Bands," presented at the 2018 IEEE International Electron Devices Meeting (IEDM), IEEE, 2018. doi: 10.1109/iedm.2018.8614564.

[4] M. Liffredo, N. Xu, S. Stettler, F. Peretti, and L. G. Villanueva, "Piezoelectric and elastic properties of Al0.60Sc0.40N thin films deposited on patterned metal electrodes," *J. Vac. Sci. Technol. A*, vol. 42, no. 4, p. 043404, Jul. 2024, doi: 10.1116/6.0003497.

[5] J. B. Shealy *et al.*, "Single Crystal AlScN-on-Silicon XBAW RF Filter Technology for Wide Bandwidth, High Frequency 5G and WiFi Applications," presented at the CS MANTECH Conference, 2022, pp. 9–12. [Online]. Available: https://akoustis.com/wp-content/uploads/2022/06/Single-Crystal-AlScN-on-Silicon-XBAW%E2%84%A2RF-Filter-Technology-for-Wide-Bandwidth-High-Frequency-5G-and-Wi-Fi-Applications.pdf

[6] A. Tag *et al.*, "Next Generation Of BAW: The New Benchmark for RF Acoustic Technologies," in *2022 IEEE International Ultrasonics Symposium (IUS)*, Venice, Italy: IEEE, Oct. 2022, pp. 1–4. doi: 10.1109/IUS54386.2022.9958625.

[7]     A. Bogner *et al.*, "Enhanced Piezoelectric Al1-xScxN RF- MEMS Resonators For Sub6 GHZ RF- Filter Applications: Design, Fabrication and Characterization".

[8]     A. Hagelauer *et al.*, "From Microwave Acoustic Filters to Millimeter-Wave Operation and New Applications," *IEEE J. Microw.*, vol. 3, no. 1, pp. 484–508, Jan. 2023, doi: 10.1109/JMW.2022.3226415.

[9]     T. Takai *et al.*, "I.H.P. SAW Technology and its Application to Microacoustic Components," in *2017 IEEE International Ultrasonics Symposium (IUS)*, 2017.

[10]    S. Inoue and M. Solal, "LT/Quartz Layered SAW Substrate with Suppressed Transverse Mode Generation," in *2020 IEEE International Ultrasonics Symposium (IUS)*, Las Vegas, NV, USA: IEEE, Sep. 2020, pp. 1–4. doi: 10.1109/IUS46767.2020.9251459.

[11]    P. Liu *et al.*, "A Spurious-Free SAW Resonator with Near-Zero TCF Using $LiNbO_3$/$SiO_2$/quartz," *IEEE Electron Device Lett.*, pp. 1–1, 2023, doi: 10.1109/LED.2023.3309672.

[12]    E. Butaud *et al.*, "Innovative Smart Cut$^{TM}$ Piezo On Insulator (POI) Substrates for 5G acoustic filters," in *2020 IEEE International Electron Devices Meeting (IEDM)*, San Francisco, CA, USA: IEEE, Dec. 2020, p. 34.6.1-34.6.4. doi: 10.1109/IEDM13553.2020.9372020.

[13]    T.-H. Hsu, K.-J. Tseng, and M.-H. Li, "Large Coupling Acoustic Wave Resonators Based on $LiNbO_3$/$SiO_2$/Si Functional Substrate," *IEEE Electron Device Lett.*, vol. 41, no. 12, pp. 1825–1828, Dec. 2020, doi: 10.1109/LED.2020.3030797.

[14]    H. Zhou *et al.*, "Ultrawide-Band SAW Devices Using SH0 Mode Wave with Increased Velocity for 5G Front-Ends," in *IEEE International Ultrasonics Symposium, IUS*, IEEE Computer Society, 2021. doi: 10.1109/IUS52206.2021.9593689.

[15]    S. Tanaka, Y. Guo, and M. Kadota, "Evolution of SAW and BAW Devices Using Thin $LiTaO_3$ and $LiNbO_3$," in *2024 IEEE MTT-S International Conference on Microwave Acoustics & Mechanics (IC-MAM)*, Chengdu, China: IEEE, May 2024, pp. 169–172. doi: 10.1109/IC-MAM60575.2024.10538491.

[16]    M. Kadota and S. Tanaka, "Wideband acoustic wave resonators composed of hetero acoustic layer structure," *Jpn. J. Appl. Phys.*, vol. 57, no. 7, Jul. 2018, doi: 10.7567/JJAP.57.07LD12.

[17]    H. Xu *et al.*, "SAW Filters on $LiNbO_3$/SiC Heterostructure for 5G n77 and n78 Band Applications," *IEEE Trans. Ultrason. Ferroelectr. Freq. Control*, vol. 70, no. 9, pp. 1157–1169, Sep. 2023, doi: 10.1109/TUFFC.2023.3299635.

[18]    P. Zheng *et al.*, "Near 5-GHz Longitudinal Leaky Surface Acoustic Wave Devices on LiNbO3/SiC Substrates," *IEEE Trans. Microw. Theory Tech.*, pp. 1–9, 2023, doi: 10.1109/TMTT.2023.3305078.

[19]    R. Su *et al.*, "Over GHz bandwidth SAW filter based on 32° Y-X LN/$SiO_2$/poly-Si/Si heterostructure with multilayer electrode modulation," *Appl. Phys. Lett.*, vol. 120, no. 25, p. 253501, Jun. 2022, doi: 10.1063/5.0092767.

[20]    L. Colombo, A. Kochhar, G. Vidal-Alvarez, P. Simeoni, U. Soysal, and G. Piazza, "Sub-GHz X-Cut Lithium Niobate $S_0$ Mode MEMS Resonators," *J. Microelectromechanical Syst.*, vol. 31, no. 6, pp. 888–900, Dec. 2022, doi: 10.1109/JMEMS.2022.3204449.

[21]    R. Tetro, L. Colombo, W. Gubinelli, G. Giribaldi, and M. Rinaldi, "X-Cut Lithium Niobate $S_0$ Mode Resonators for

5G Applications," in *2024 IEEE 37th International Conference on Micro Electro Mechanical Systems (MEMS)*, Austin, TX, USA: IEEE, Jan. 2024, pp. 1102–1105. doi: 10.1109/MEMS58180.2024.10439491.

[22] M. Faizan and L. G. Villanueva, "Frequency-scalable fabrication process flow for lithium niobate based Lamb wave resonators," *J. Micromechanics Microengineering*, vol. 30, no. 1, 2020, doi: 10.1088/1361-6439/ab5b7b.

[23] G. Giribaldi, L. Colombo, and M. Rinaldi, "6-20 GHz 30% ScAlN Lateral Field-Excited Cross-sectional Lame' Mode Resonators for future mobile RF Front-Ends," *IEEE Trans. Ultrason. Ferroelectr. Freq. Control*, pp. 1–1, 2023, doi: 10.1109/TUFFC.2023.3312913.

[24] G. Giribaldi, L. Colombo, P. Simeoni, and M. Rinaldi, "Compact and wideband nanoacoustic pass-band filters for future 5G and 6G cellular radios," *Nat. Commun.*, vol. 15, no. 1, p. 304, Jan. 2024, doi: 10.1038/s41467-023-44038-9.

[25] M. Kadota and T. Ogami, "5.4 GHz Lamb Wave Resonator on LiNbO3 Thin Crystal Plate and Its Application," *Jpn. J. Appl. Phys.*, vol. 50, no. 7S, p. 07HD11, Jul. 2011, doi: 10.1143/JJAP.50.07HD11.

[26] R. Lu, Y. Yang, S. Link, and S. Gong, "A1 Resonators in 128 deg Y-cut Lithium Niobate with Electromechanical Coupling of 46.4%," *J. Microelectromechanical Syst.*, 2020, doi: 10.1109/JMEMS.2020.2982775.

[27] V. Plessky, S. Yandrapalli, P. J. Turner, L. G. Villanueva, and M. Faizan, "Laterally excited bulk wave resonators (XBARs) based on thin Lithium Niobate platelet for 5GHz and 13 GHz filters".

[28] M. Kadota and S. Tanaka, "Ultra-wideband ladder filters using zero-th shear mode plate wave in ultrathin LiNbO3 plate with apodized interdigital transducers," in *Japanese Journal of Applied Physics*, Japan Society of Applied Physics, Jul. 2016. doi: 10.7567/JJAP.55.07KD04.

[29] J. Zou *et al.*, "Ultra-Large-Coupling and Spurious-Free SH0 Plate Acoustic Wave Resonators Based on Thin LiNbO3," *IEEE Trans. Ultrason. Ferroelectr. Freq. Control*, vol. 67, no. 2, pp. 374–386, Feb. 2020, doi: 10.1109/TUFFC.2019.2944302.

[30] S. Wu *et al.*, "Large Coupling and Spurious-Free SH0 Plate Acoustic Wave Resonators Using LiNbO3 Thin Film," *IEEE Trans. Electron Devices*, pp. 1–8, 2023, doi: 10.1109/TED.2023.3297561.

[31] H. Bhugra and G. Piazza, *Piezoelectric MEMS Resonators*. 2017. doi: 10.1007/978-3-319-28688-4.

[32] P. J. Turner *et al.*, "5 GHz Band n79 wideband microacoustic filter using thin lithium niobate membrane," *Electron. Lett.*, vol. 55, no. 17, pp. 942–944, Aug. 2019, doi: 10.1049/el.2019.1658.

[33] S. Yandrapalli, S. E. K. Eroglu, V. Plessky, H. B. Atakan, and L. G. Villanueva, "Study of Thin Film LiNbO3 Laterally Excited Bulk Acoustic Resonators," *J. Microelectromechanical Syst.*, vol. 31, no. 2, pp. 217–225, Apr. 2022, doi: 10.1109/JMEMS.2022.3143354.

[34] O. Barrera, S. Cho, J. Kramer, V. Chulukhadze, J. Campbell, and R. Lu, "38.7 GHz Thin Film Lithium Niobate Acoustic Filter," in *2024 IEEE International Microwave Filter Workshop (IMFW)*, Cocoa Beach, FL, USA: IEEE, Feb. 2024, pp. 87–90. doi: 10.1109/IMFW59690.2024.10477121.

[35] F. Hartmann, S. E. Kuçuk Eroglu, E. Navarro-Gessé, C. Collado, J. Mateu, and L. G. Villanueva, "A 5G n77 Filter Using Shear Bulk Mode Resonator With Crystalline X-cut Lithium Niobate Films," in *2024 IEEE International Microwave*

*Filter Workshop (IMFW)*, Cocoa Beach, FL, USA: IEEE, Feb. 2024, pp. 78–80. doi: 10.1109/IMFW59690.2024.10477114.

[36] S. Yandrapalli, S. K. Eroglu, J. Mateu, C. Collado, V. Plessky, and L. G. Villanueva, "Toward Band n78 Shear Bulk Acoustic Resonators Using Crystalline Y-Cut Lithium Niobate Films With Spurious Suppression," *J. Microelectromechanical Syst.*, vol. 32, no. 4, pp. 327–334, Aug. 2023, doi: 10.1109/JMEMS.2023.3282024.

[37] S. Stettler and L. G. Villanueva, "Transversal Spurious Mode Suppression in Ultra-Large-Coupling SH0 Acoustic Resonators on YX36°-Cut Lithium Niobate," *J. Microelectromechanical Syst.*, vol. 32, no. 3, pp. 279–289, Jun. 2023, doi: 10.1109/JMEMS.2023.3262021.

[38] R. Tetro, L. Colombo, and M. Rinaldi, "2–16 GHz Multifrequency X-Cut Lithium Niobate NEMS Resonators on a Single Chip," Jun. 12, 2024, *arXiv*: 2406.07784. doi: 10.48550/ARXIV.2405.05547.

[39] V. Chulukhadze *et al.*, "2 to 16 GHz Fundamental Symmetric Mode Acoustic Resonators in Piezoelectric Thin-Film Lithium Niobate," May 13, 2024, *arXiv*: 2405.08139. doi: 10.48550/ARXIV.2405.08139.

[40] M. Kadota, T. Kojima, and S. Tanaka, "2–8 GHz Range High Harmonic SAW Resonator with Grooved Electrodes in LiNbO$_3$," in *2021 IEEE International Ultrasonics Symposium (IUS)*, Xi'an, China: IEEE, Sep. 2021, pp. 1–4. doi: 10.1109/IUS52206.2021.9593579.

[41] A. Clairet *et al.*, "Electrode Confined Acoustic Wave (ECAW) devices for Ultra High Band applications," in *2023 IEEE International Ultrasonics Symposium (IUS)*, Montreal, QC, Canada: IEEE, Sep. 2023, pp. 1–6. doi: 10.1109/IUS51837.2023.10306791.

[42] M. Kadota, T. Kimura, and Y. Ida, "Tunable Filters Using Ultrawide-Band Surface Acoustic Wave Resonator Composed of Grooved Cu Electrode on LiNbO$_3$," *Jpn. J. Appl. Phys.*, vol. 49, no. 7S, p. 07HD26, Jul. 2010, doi: 10.1143/JJAP.49.07HD26.

[43] V. Plessky and Koskela, Julius, "Coupling-of-modes analysis of SAW devices," *Int. J. High Speed Electron. Syst.*, vol. 10, no. 04, pp. 867–947, Dec. 2000.

[44] F. B. Kaufman *et al.*, "Chemical-Mechanical Polishing for Fabricating Patterned W Metal Features as Chip Interconnects," *J. Electrochem. Soc.*, vol. 138, no. 11, pp. 3460–3465, Nov. 1991, doi: 10.1149/1.2085434.

[45] Oliver, Michael R., *Chemical-Mechanical Planarization of Semiconductor Materials*, 1st ed., vol. 69. in Springer Series in Materials Science, vol. 69. Heidelberg: Springer Berlin, 2004.

[46] J.-B. Yoon, G. Y. Oh, C.-H. Han, E. Yoon, and C.-K. Kim, "Planarization and trench filling on severe surface topography with thick photoresist for MEMS," presented at the Micromachining and Microfabrication, J. H. Smith, Ed., Santa Clara, CA, Aug. 1998, pp. 297–306. doi: 10.1117/12.324314.

[47] "IEEE Standard on Piezoelectricity Standards Committee of the IEEE Ultrasonics, Ferroelectrics, and Frequency Control Society IEEE Standards Board American National Standards Institute," 1987.

# Acknowledgments

S.S. and L.G.V. acknowledge support from the Swiss National Science Foundation (SNSF) under projects CRSII5_189967 and 200020_184935. S.S. thanks the staff of the EPFL Center of MicroNanoTechnology (CMi) for their support with the fabrication process and tool maintenance.

# Ethics declarations

## Competing interests

The authors declare the following competing interests: A patent application related to this technology has been filed by EPFL with S.S. and L.G.V. as inventors.

# Author contributions

**Silvan Stettler:** Conceptualization (equal); Data curation (lead); Formal Analysis (lead); Investigation (lead); Methodology (lead); Software (lead); Validation (lead); Visualization (lead); Writing – original draft (lead); Writing – review & editing (lead). **Guillermo Villanueva:** Conceptualization (Lead); Funding acquisition (Lead); Project administration (Lead); Resources (Lead); Supervision (Lead); Writing – review & editing (Equal).

# Data availability statement

The data that support the findings of this study are available from the corresponding author upon reasonable request.

# Supplementary Information: *Suspended lithium niobate acoustic resonators with buried electrodes for radiofrequency filtering*

Silvan Stettler [1], Luis Guillermo Villanueva [1, †]

[1] *Institute of Mechanical Engineering, École Polytechnique Fédérale de Lausanne (EPFL), CH-1015 Lausanne, Switzerland*

## Contents

† guillermo.villanueva@epfl.ch

## S1: Piezoelectric tensor of YX36°-cut LiNbO₃

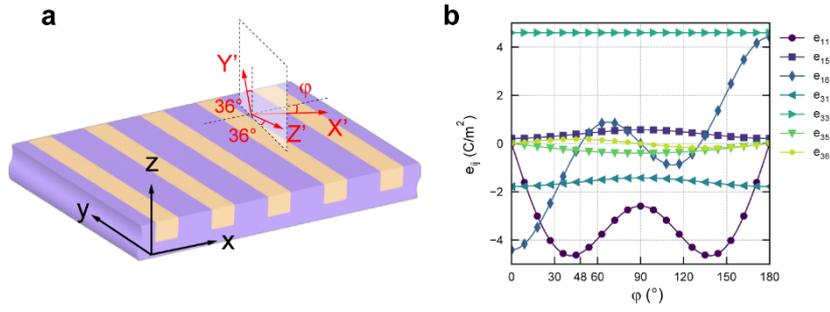

**Fig. S1 a,** Orientation of the crystalline axis (in red, X'-Y'-Z') for YX36°-cut LiNbO₃ with respect to a reference coordinate system aligned with the transducer (in black, x-y-z). $\varphi$ designates the in-plane orientation of the transducer on the chip. **b,** Components of the piezoelectric tensor for YX36°-cut LiNbO₃ expressed in the reference coordinate system for varying $\varphi$. At $\varphi = 0°$, $e_{16}$ is maximum and $e_{11}$ is zero. At $\varphi = 48°$, $e_{11}$ is close to maximum and $e_{16}$ is zero. Piezoelectric properties for LiNbO₃ were taken from [1].

## S2: Simulated admittance and mode shapes in periodic 2D transducer unit cells

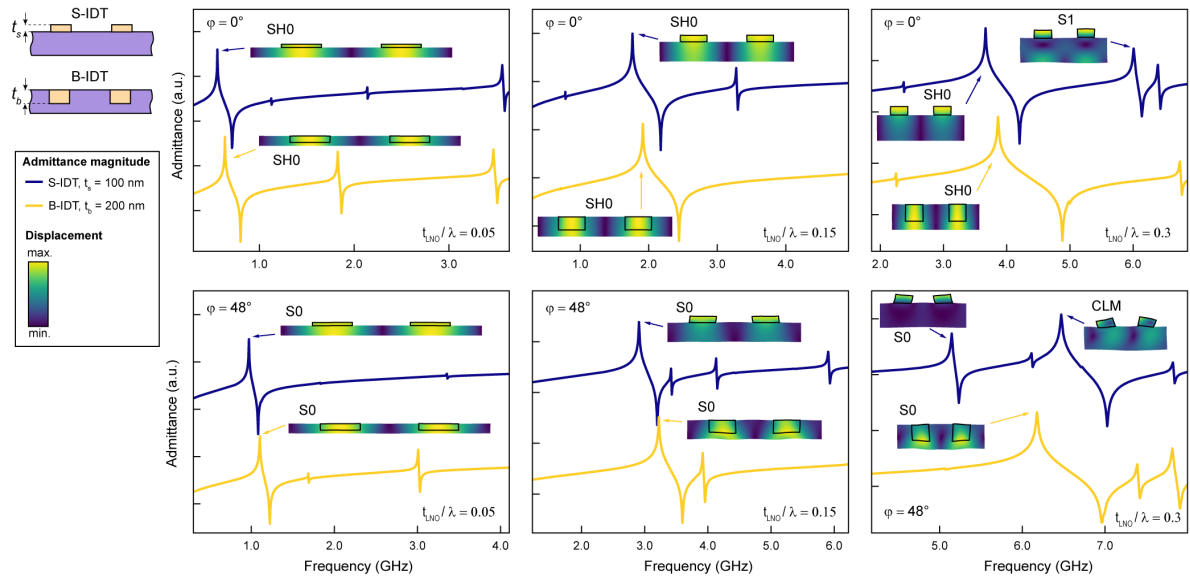

**Fig. S2** FEM-simulated admittance and mode shapes for $\varphi = 0°, 48°$ for varying $t_{LNO}/\lambda$ with B-IDT and S-IDT architectures. A 2D model of a unit cell of the transducer with periodic boundary conditions is used for the simulation.

## S3: Propagation of SH0 and S0 acoustic waves in periodic B-IDT gratings

Fig. S3 shows the simulated dispersion relation of SH0 and S0 waves propagating in a B-IDT grating with an electrode pitch ($p = \lambda/2$) of 550 nm. A 2D model of a unit cell of the transducer (including electrodes with sidewalls at a 70° angle) with periodic boundary conditions is used for the simulation. The dispersion relations were obtained by sweeping the complex wavenumber ($k = q' + iq''$) of the periodic boundary conditions. $k_0 = \pi/p$ is the fundamental wavenumber of the grating.

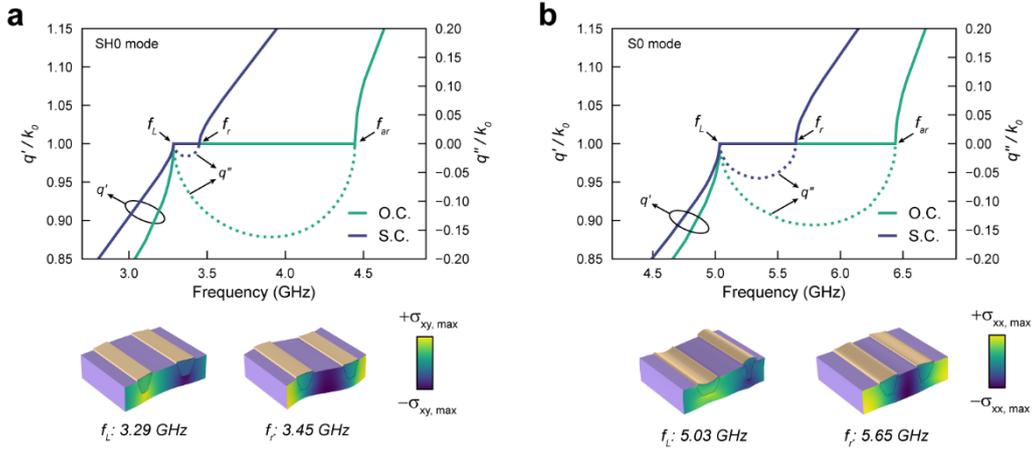

**Fig. S3 a, b,** Simulated dispersion in the B-IDT grating of the SH0 (a) and S0 (b) modes for $p$ = 550 nm with the electrodes in short-circuit (S.C.) and open-circuit (O.C.) conditions. In the stopband, $q'$ is equal to the fundamental wavenumber of the grating $k_0$. $f_r$ and $f_{ar}$ appear at the upper stopband edges in S.C. and O.C. conditions, respectively. The lower stopband edges ($f_L$) correspond to non-excited mode with maxima of stress ($\sigma_{xy}$, $\sigma_{xx}$) located under the electrodes.

## S4: Effect of edge reflector placement on admittance response

Fig. S4a shows the admittance of three identical SH0 resonators with edge reflectors fabricated on the same chip but at different locations on the chip. The response in terms of spurious mode frequency and amplitude to the left of the main resonance is notably different for all three devices. For these resonators, the transducer consists of $N_p = 70$ electrode pairs. The optimal placement of the edge reflector, (i.e. the free edge created by etching the release holes) is in the center of the last electrode on both sides of the transducer. This choice coincides with the nodal points of the stress fields of the generated acoustic wave. Hence, any reflected wave from either side is in phase with the wave generated by the transducer resulting in a resonating standing wave [2]. However, the placement of the edge reflector relies on accurate lithographic alignment of the mask to pattern the release holes with respect to the B-IDT grating. With the alignment scheme and chip-level fabrication used for this work, misalignment up to 200 nm is possible and is found to vary across the chip. Using a finite 2D FEM model, the impact of misalignment ($\Delta$) on the measured admittance response around the main resonance frequency for SH0 and S0 is simulated and shown in Fig. S4b. The simulation results show that for $\Delta$ on the order of 100 nm, the admittance features a significant ripple to the left of the main resonance like the measured response. Further, the frequency of this spurious mode is very sensitive to Δ which explains the variance observed across the sample (Fig. S4a).

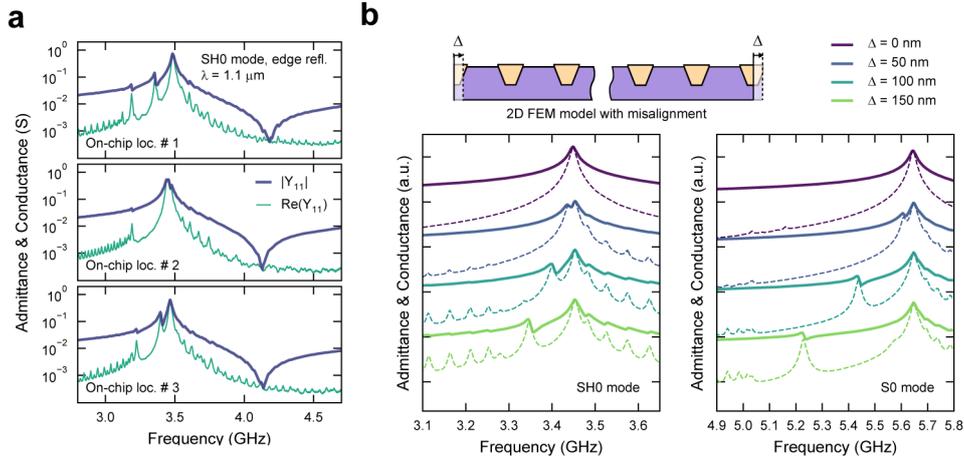

**Fig. S4 a,** Measured admittance of three identical devices (with edge reflectors), placed at different locations on the same chip. **b,** FEM-simulated admittance close to resonance using a 2D finite model with $\lambda = 1.1\ \mu m$ and $N_p = 70$ electrode pairs including the free edges at the B-IDT ends. The geometry of the buried electrodes includes a sloped sidewall to replicate the fabricated structure as closely as possible. The position of the free edges is varied with respect to its ideal position to mimic lithographic misalignment (Δ). $\Delta = 0\ nm$ corresponds to edges placed exactly in the center of the last electrode.

## S5: Process flow for B-IDT resonator fabrication

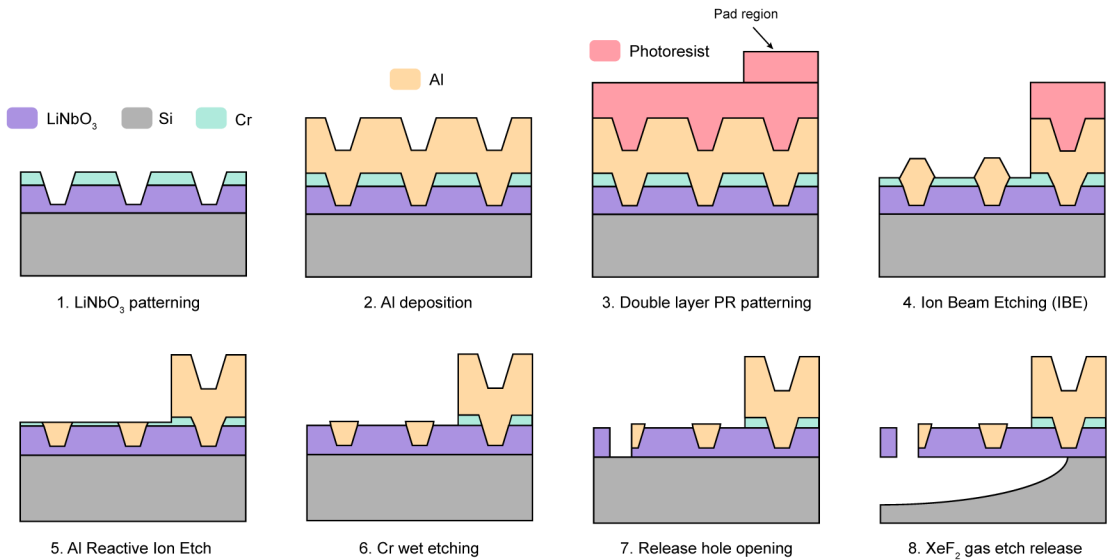

**Fig. S5** Schematic overview of the processing steps used for B-IDT resonator fabrication.

# Supplementary References

[1] G. Kovacs, M. Anhorn, H. E. Engan, G. Visintini, and C. C. W. Ruppel, "Improved material constants for LiNbO$_3$ and LiTaO$_3$", presented at the IEEE Symposium on Ultrasonics, IEEE, Dec. 2002, pp. 435–438. doi: 10.1109/ultsym.1990.171403

[2] J. Zou *et al.*, "Ultra-Large-Coupling and Spurious-Free SH0 Plate Acoustic Wave Resonators Based on Thin LiNbO3," *IEEE Trans. Ultrason. Ferroelectr. Freq. Control*, vol. 67, no. 2, pp. 374–386, Feb. 2020, doi: 10.1109/TUFFC.2019.2944302